\documentstyle[12pt,epsf]{article}
\catcode`\@=11
\def\slash{\mathpalette\make@slash}
\def\make@slash#1#2{\setbox\z@\hbox{$#1#2$}%
  \hbox to 0pt{\hss$#1/$\hss\kern-\wd0}\box0}
\catcode`\@=12 

\begin{document}

\begin{center}
{\large \bf $B_c$-meson sum rules at next-to-leading order}\\
\vspace*{5mm} Andrei I. Onishchenko

\end{center}
\begin{center}
{\it Institute for Theoretical and Experimental Physics,\\
B. Cheremushkinskaja 25, Moscow, 117259 Russia}
\end{center}

\begin{abstract}
{We present $B_c$-meson two-point sum rules at
next-to-leading order in NRQCD approximation. Analitycal results
for perturbative spectral density and gluon condensate
contribution with account for summed Coulomb corrections are
given. Estimates of various $c$-quark  masses together with
couplings and masses of lowest lying $B_c$-meson resonances are
performed. }
\end{abstract}

\section{Introduction}

The study of the physics of $B_c$-meson\footnote{For review see
\cite{review}.} has already a long story. The collaborative
efforts of many scientists around the world have led to remarkable
predictions, describing spectroscopy \cite{eichten,prd1}, decays
\cite{pm,beneke,Col,bagan,Kis1} and production mechanisms
\cite{prod} of this object. The theoretical work done was very
helpful in the experimental search and discovery of this
meson by CDF collaboration \cite{cdf}. As it was many times
earlier with other particles, the observation of $B_c$-meson in
nature have not diminished the interest of physicists to this
object. People wonder how much we can learn from this meson about
the Standard Model (SM) of particle interactions. In this paper we
explore a potential of $B_c$ - meson in extracting the parameters
of SM lagragian, namely, the heavy quark masses.

What concerns its spectroscopic properties, this meson stands
among the families of charmonium $\bar c c$ and bottominum $\bar b
b$ : two heavy quarks move nonrelativistically, since the
confinement scale, determining the presence of light degrees of
freedom (sea of gluons and quarks), is suppressed with respect to
the heavy quark masses $m_Q$ as well as the Coulomb-like exchanges
result in transfers about $\alpha_s m_Q^2$, which is again much
less than the heavy quark mass. It is precisely the
nonrelativistic nature of heavy quark dynamics in the $B_c$-meson,
that offers us a possibility to use the NRQCD framework and gain
more inside on QCD dynamics of the constituent heavy quarks.
Recently, within this framework the NLO and NNLO NRQCD sum rules
were derived and analyzed for the case of $\Upsilon$-family
\cite{NNLO}. The result of this analysis was a precise
determination of pole, running, 1S, potential subtracted and
kinetic $b$-quark masses. Here we employ the same NRQCD sum rule
method to determine the numerical values for different definitions
of $c$-quark masses together with couplings and masses of lowest
lying $B_c$-meson resonances. The present work is similar to the
analysis performed for $\Upsilon$-family with differences in
particular analytical expressions for theoretical moments of
two-point correlator.

The paper is organized as follows. In section 2 we remind the reader
the QCD sum rule framework for the two-point sum rule. Section 3
introduces the NRQCD framework and we comment on its connection
with QCD one. Section 4 explains the strategy used for
calculation of NRQCD two-point correlation function. In section 5
we calculate the perturbative theoretical expressions for moments
of correlation function. Section 5 deals with the corrections
to the correlator, coming from the gluon condensate operator. In
section 7 we present our results on the different $c$-quark mass
definitions, $B_c$-meson mass and coupling constant. The detail
discussion on optimization methods is also given. And
finally, section 8 contains our summary.

\section{Two-point sum rules}

In the QCD sum rules approach\cite{Shif,Rein} meson bound states
are described by local interpolating currents of the form $J =
\bar q_1\Gamma q_2$, where $\Gamma$ is a suitable Dirac structure
to account for meson quantum numbers. Thus, the studied object in
QCD sum rules is a coupling of meson under consideration to
corresponding current. In the case of $B_c$ meson this coupling is
defined by the following equation
\begin{equation}
\langle 0|\bar b i\gamma_5 c|B_c\rangle =
\frac{f_{B_c}M_{B_c}^2}{m_b + m_c}
\end{equation}
The estimates of $B_c$-meson structure constant in QCD sum rule
framework were already performed in a number of papers
\cite{Valera1}. The distinctive feature of the present work is a
complete next-to-leading order analysis, containing correct
treatment of Coulomb corrections. For discussion of importance of
such corrections see, for example \cite{Kis1}.

In QCD sum rules framework, the $B_c$-meson structure constant is
naturally obtained from two-point correlator of the following form
\begin{equation}
\Pi (q^2) = i\int d^4 x e^{i q\cdot x}\langle 0|
J_{B_c}(x)J_{B_c}(0)^{+}|0\rangle, \label{Cor1}
\end{equation}
where $J_{B_c} = \bar b i\gamma_5 c$.

The left hand side of Eq.(\ref{Cor1}) can be computed in QCD, for
$|q^2|$ much larger than $\Lambda^{2}_{QCD}$ (or , what
alternatively, for $|q^2 - (m_c+m_b)^2|$ much larger than
$\Lambda^{2}_{QCD}$), with the use of short-distance Operator
Product Expansion (OPE) for correlation function under
consideration.
\begin{equation}
\Pi (q^2)_{QCD} = \Pi_{pert}(q^2) +
C_{G^2}(q^2)\langle\frac{\alpha_s}{\pi}G^2\rangle + \ldots\quad
,\label{OPE}
\end{equation}
where $C_{G^2}(q^2)$ is a Wilson coefficient of gluon condensate
operator and dots in right hand side of Eq.(\ref{OPE}) present
contributions of operators with higher dimension $(d > 4)$.

The connection to physical spectrum of $B_c$-meson can be obtained
by writing the following dispersion relation
\begin{equation}
\Pi (q^2) = \frac{1}{\pi}\int\frac{\rho (s)_{had} d s}{s - q^2} +
\mbox{subtractions},
\end{equation}
where
\begin{eqnarray}
\rho (s)_{had} = \pi\frac{f_{B_c}^2M_{B_c}^4}{(m_b + m_c)^2}\delta
(s - M_{B_c}^2) + \rho (s)_{pert}\theta (s - s_{thr}),
\label{rhoh}
\end{eqnarray}
Here we have taken into account only lowest lying $B_c$-meson
state, $s_{thr}$ is a continuum threshold and $\rho (s)_{pert}$ is
connected to $\Pi_{pert}(q^2)$ via the following dispersion
relation
\begin{equation}
\Pi_{pert} (q^2) = \frac{1}{\pi}\int\frac{\rho (s)_{pert} d s}{s -
q^2} + \mbox{subtractions}
\end{equation}
In the numerical analysis we however explore a different anzaz, on
which we will comment in the section with numerical results.

There are several schemes of QCD sum rules, which can be used for
the extraction of quantities, you are interesting in. The most
popular among them are momentum and Borel schemes. In this paper
we will employ the first one and the studied object will be the
momentum of two-point correlation function, given by the following
expression
\begin{eqnarray}
P_n = n!\left(\frac{d}{dq^2}\right)^n\Pi (q^2)|_{q^2 = 0}
\end{eqnarray}

Thus far we have discussed QCD sum rule framework for
determination of $B_c$-meson structure constant. As a primary goal
of this paper is to perform a consistent analysis of the same
quantity in NRQCD approximation, in the next section we define
NRQCD sum rule framework and comment on connection of the latter
with QCD sum rule analysis.

\section{NRQCD approximation}

In this section we set up a consistent NRQCD framework, in which
the two-point correlation function $\Pi (q^2)$ can be determined
in a systematic manner at next-to-leading order. Our presentation
in this and the next sections closely follow that of
\cite{Hoang2}, so we advice the reader to read that paper for more
detail.

NRQCD is an effective field theory of QCD designed to handle
nonrelativistic heavy-quark-antiquark systems to in principle
arbitrary precision. Considering all quarks of the first and
second generation as massless the NRQCD Lagrangian
reads~\cite{Bodwin1}
\begin{eqnarray}
\lefteqn{ {\cal{L}}_{\mbox{\tiny NRQCD}} \, = \, - \frac{1}{2}
\,\mbox{Tr} \, G^{\mu\nu} G_{\mu\nu} + \sum_{q=u,d,s,c} \bar q \,
i \slash{D} \, q }\nonumber\\[2mm] & & +\, \psi^\dagger\,\bigg[\,
i D_t + a_1\,\frac{{\mbox{\boldmath $D$}}^2}{2\,M_t} +
a_2\,\frac{{\mbox{\boldmath $D$}}^4}{8\,M_t^3} \,\bigg]\,\psi +
\ldots \nonumber\\[2mm] & & + \,\psi^\dagger\,\bigg[\,
\frac{a_3\,g}{2\,M_t}\,{\mbox{\boldmath $\sigma$}}\cdot
    {\mbox{\boldmath $B$}}
+ \, \frac{a_4\,g}{8\,M_t^2}\,(\,{\mbox{\boldmath $D$}}\cdot
  {\mbox{\boldmath $E$}}-{\mbox{\boldmath $E$}}\cdot
  {\mbox{\boldmath $D$}}\,)
+ \frac{a_5\,g}{8\,M_t^2}\,i\,{\mbox{\boldmath $\sigma$}}\,
  (\,{\mbox{\boldmath $D$}}\times
  {\mbox{\boldmath $E$}}-{\mbox{\boldmath $E$}}\times
  {\mbox{\boldmath $D$}}\,)
 \,\bigg]\,\psi
+\ldots \nonumber\\[2mm] & & + \mbox{$\chi\chi^\dagger$ bilinear
terms and higher dimensional operators} \,. \label{NRQCDLagr}
\end{eqnarray}
The gluons and massless quarks are described by the conventional
relativistic Lagrangian, where $G_{\mu\nu}$ is the gluon field
strength tensor, $q$ the Dirac spinor of a massless quark and
$D_\mu$ the gauge covariant derivative. For convenience, all color
indices in Eq.~(\ref{NRQCDLagr}) and throughout this work are
suppressed. The nonrelativistic $c$ and $\bar b$ quarks are
described by the Pauli spinors $\psi$ and $\chi$, respectively.
$D_t$ and {\boldmath $D$} are the time and space components of the
gauge covariant derivative $D$ and $E^i = G^{0 i}$ and $B^i =
\frac{1}{2}\epsilon^{i j k} G^{j k}$ the electric and magnetic
components of the gluon field strength tensor (in Coulomb gauge).
The straightforward $\chi^\dagger \chi$ bilinear terms are omitted
and can be readily obtained. The short-distance coefficients
$a_1,\ldots,a_5$ are normalized to one at the Born level. The
actual form of the higher order contributions to the
short-distance coefficients $a_1,\ldots,a_5$ is irrelevant for this work,
as we will later use the ``direct matching'' procedure at the
level of the final result for the correlation function.

Now let us discuss our correlation function in NRQCD
approximation. Eq.(\ref{Cor1}) can be rewritten as
\begin{equation}
\Pi (q^2) = i\langle
0|T\tilde{J}_{B_c}(q)\tilde{J}_{B_c}(-q)^{+}|0\rangle,
\label{Cor2}
\end{equation}
where $\tilde{J}_{B_c}(q) = (\bar b i\gamma_5 c) (q)$. Expressing
Dirac fields for $\bar b$ and $c$ - quarks in terms of Pauli
spinors $\chi$ and $\psi$
\begin{eqnarray}
u_c(\mbox{{\bf q}}) &=& \sqrt{E_c+m_c \over 2E_c} \left(
\begin{array}{c}
    \psi \\
    {\mbox{{\bf q}} \cdot \mbox{{\boldmath $\sigma$}}
        \over E_c+m_c} \psi
\end{array}
    \right) ,
\label{uspinor}
\\
v_b(-\mbox{{\bf q}}) &=& \sqrt{E_b+m_b \over 2E_b}    \left(
\begin{array}{c}
    {(-\mbox{{\bf q}}) \cdot \mbox{{\boldmath $\sigma$}}
        \over E_b+m_b} \chi \\
    \chi
\end{array}
    \right) ,
\label{vspinor}
\end{eqnarray}
we have\footnote{Here and later in the paper, except stated
otherwise, by $m_c$ and $m_b$ we mean the pole heavy quark
masses.}
\begin{eqnarray}
\tilde{J}_{B_c}(q) &\;\approx\;&
- i\left((\chi^\dagger \psi)(q) \;-\; {1 \over 8} \left( {m_b - m_c
\over m_b m_c} \right)^2 (({\bf D} \chi)^\dagger
\cdot {\bf D} \psi)(q) \;+\; \ldots\right) \\
\tilde{J}_{B_c}(-q)^{\dagger} &\;\approx\;&
- i\left((\psi^\dagger \chi)(-q) \;-\; {1 \over 8} \left( {m_b - m_c
\over m_b m_c} \right)^2 (({\bf D} \psi)^\dagger
\cdot {\bf D} \chi)(-q) \;+\; \ldots\right)
\end{eqnarray}
Inserting these expansions into Eq. (\ref{Cor2}) we obtain
\begin{eqnarray}
i \Pi (q^2) &=&  C_1 (\mu_{\rm hard},\mu_{\rm fact}){\cal A}_1 (E,\mu_{\rm
soft},\mu_{\rm fact})
\nonumber\\
&& - {1 \over 4} \left( {m_b - m_c
\over m_b m_c} \right)^2 C_2 (\mu_{\rm hard},\mu_{\rm fact}){\cal A}_2
(E,\mu_{\rm soft},\mu_{\rm fact}) + \ldots , \label{Match}
\end{eqnarray}
where
\begin{eqnarray}
{\cal A}_1 &=& \langle 0|(\chi^{\dagger}\psi )(\psi^{\dagger}\chi )|0\rangle,
\\
{\cal A}_2 &=& \frac{1}{2}\langle 0|(\chi^{\dagger}\psi )
(({\bf D} \psi)^\dagger\cdot {\bf D} \chi) + \mbox{h.c.}|0\rangle,
\end{eqnarray}
The right-hand side of Eq. (\ref{Match}) just represents an
application of the factorization formalism proposed
in~\cite{Bodwin1}. The second term in this expression is
suppressed by $v^2$, i.e.\ of next-to-next-to-leading order and
thus of no relevance to us in present analysis. The
nonrelativistic current correlator ${\cal{A}}_1$ contains the
resummation of the singular Coulomb terms. It incorporates all the
long-distance, dynamics governed by soft scales like the relative
three momentum $\sim m_{red}v$ or the binding energy of the $c\bar
b$ system $\sim m_{red}v^2$. The constant $C_1$ (it is normalized
to one at the Born level), on the other hand, describes
short-distance effects involving hard scales of the order of heavy
quark mass. It is represented only by a simple power series in
$\alpha_s$ and does not contain any resummations in $\alpha_s$. We
would also like to note, that $C_1$ is independent of $q^2$. In
Eq.~(\ref{Match}) we have also indicated the dependence of the
NRQCD correlators and the short-distance coefficients on the
various renormalization scales: The factorization scale $\mu_{\rm
fact}$ essentially represents the boundary between hard and soft
momenta. The dependence on the factorization scale becomes
explicit because of ultraviolet (UV) divergences contained in
NRQCD. Because, as in any effective field theory, this boundary is
not defined unambiguously, both the correlators and the
short-distance coefficients in general depend on $\mu_{\rm fact}$.
The soft scale $\mu_{\rm soft}$ and the hard scale $\mu_{\rm
hard}$, on the other hand, are inherent to the correlators and the
short-distance constants, respectively, governing their
perturbative expansion. If we would have all orders in $\alpha_s$
and $v$ at hand, the dependence of correlation function $\Pi
(q^2)$ on variations of each the three scales would vanish
exactly. Unfortunately, we only perform the calculation up to
next-to-leading order in $\alpha_s$ and $v$ which leads to a
residual dependence\footnote{Here we would like to note, that the
object studied $P_n$ at NLO does not depend on $\mu_{fact}$} on
the three scales $\mu_{\rm fact}$, $\mu_{\rm soft}$ and $\mu_{\rm
hard}$.

\section{Calculation of NRQCD correlator}

To calculate the NRQCD correlator ${\cal{A}}_1$  we use methods
originally developed for QED bound state calculations in the
framework of NRQED~\cite{Hoang1,Caswell1,Lepage1,Hoang5} and apply
them to $B_c$-meson bound state description in the kinematic
regime close to the threshold. At next-to-leading order quarks,
composing $B_c$-meson experience only instantaneous interactions,
given by the following potentials
\begin{eqnarray}
V_c^{(0)}(\vec r) &=& -\frac{C_F\alpha_s}{r},\\ V_c^{(1)}(\vec r)
&=& V_c^{(0)}\left(\frac{\alpha_s}{4\pi}\right) \left[2\beta_0\ln
(\tilde{\mu}r) + a_1\right],\quad \tilde{\mu}\equiv
e^{\gamma_E}\mu_{\rm soft}, \label{V1}
\end{eqnarray}
where
\begin{eqnarray}
\beta_0 &=& \frac{11}{3}C_A - \frac{4}{3}T n_l, \nonumber\\ a_1
&=& \frac{31}{9}C_A - \frac{20}{9}T n_l, \\ n_l &=& 3. \nonumber
\end{eqnarray}
Here $r\equiv |\vec r|$, $C_F = \frac{4}{3}$, $C_A = 3$, $T = \frac{1}{2}$,
$\alpha_s\equiv\alpha_s (\mu_{\rm soft})$ and $\gamma_E$ is the
Euler-Mascheroni constant.

Thus, we can conclude that the problem of $B_c$-meson description
close to threshold can be treated as a pure quantum two-body
problem, so that we can use the well known analytic solutions of
the nonrelativistic Coulomb problem for positronium
\cite{Wichmann1,Hostler1,Schwinger1} and  Rayleigh-Schr\"odinger
time-independent perturbation theory (TIPT) to determine the
corrections caused by all higher order interactions and effects.

The calulational procedure for two-point NRQCD correlation
function may be devided in the following two steps

\vspace{2mm}\noindent\hspace{2mm} Step 1:
\begin{minipage}[t]{15cm} {\it Solution of the Schr\"odinger
equation.} -- The Green function of the next-to leading
Schr\"odinger equation is calculated incorporating the potentials
displayed above. The correlator ${\cal{A}}_1$ is directly related
to the zero-distance Green function of the Schr\"odinger equation.
\end{minipage}

\vspace{2mm}\noindent\hspace{2mm}
Step 2: \begin{minipage}[t]{15cm}
{\it Matching calculation.} -- The short-distance constant $C_1$ is
determined at ${\cal{O}}(\alpha_s)$ by matching  QCD current
$J_{B_c}$ to corresponding NRQCD one.
\end{minipage}

\subsection{Solution of the Schr\"odinger equation}

The nonrelativistic correlator ${\cal{A}}_1$ is calculated by
determining the Green function of the Schr\"odinger equation
($E\equiv \sqrt{q^2}- (m_b+mc)$)
\begin{equation}
\bigg(\, -\frac{\vec\nabla^2}{2m_{red}} + \bigg[\,
  V_{c}^{(0)}(\vec r) + V_{c}^{(1)}(\vec r)
\,\bigg]
- E\,\bigg)\,G(\vec r,\vec r^\prime, E) = \, \delta^{(3)}(\vec r-\vec r^\prime)
\,\label{Schroed}
\end{equation}
The relation between the correlator ${\cal{A}}_1$ and Green
function reads
\begin{eqnarray}
{\cal{A}}_1 & = & 6\,\Big[\, \lim_{|\vec r|,|\vec r^\prime|\to
0}\,G(\vec r,\vec r^\prime, E) \,\Big] \,. \label{A1toGF}
\end{eqnarray}
Eq.~(\ref{A1toGF}) can be quickly derived from the facts that
$G(\vec r,\vec r^\prime,\tilde E)$ describes the propagation of
$\bar b$ and $c$ quark pair, which is produced and annihilated at
relative distances $|\vec r|$ and $|\vec r^\prime|$, respectively,
and that the same quark pair is produced and annihilated through
the $J_{B_c}$ current at zero distances. Therefore ${\cal{A}}_1$
must be proportional to $\lim_{|\vec r|,|\vec r^\prime|\to
0}\,G(\vec r,\vec r^\prime, E)$. The correct proportionality
constant can then be determined by matching the result of full QCD
for perturbative spectral density , presented in Appendix A,  to
the imaginary part of the Green function of the Coulomb
nonrelativistic Schr\"odinger equation. We would like to emphasize
that the zero-distance Green function on the right hand side of
Eqs.~(\ref{A1toGF}) contains UV divergences which have to
regularized. In the actual calculations we impose the explicit
short-distance cutoff $\mu_{\rm fact}$. As mentioned before, this
is the reason why the correlators and the short-distance constants
depend explicitly on the (factorization) scale $\mu_{\rm fact}$.
In this work we solve equation~(\ref{Schroed}) perturbatively by
starting from well known Green function $G_c^{(0)}$ of the
nonrelativistic Coulomb
problem~\cite{Wichmann1,Hostler1,Schwinger1}
\begin{equation}
\bigg(\,-\frac{\nabla^2}{m_{red}} - V_c^{(0)}(\vec r)
- E\,\bigg)\,G_c(\vec r,\vec r^\prime, E)
\, = \, \delta^{(3)}(\vec r-\vec r^\prime)
\label{Schroed0}
\end{equation}
and incorporate all the higher order terms using TIPT.

The most general form of the Coulomb Green function reads
($r\equiv |\vec r|$, $r^\prime\equiv |\vec r^\prime|$)
\begin{eqnarray}
\lefteqn{
G_c^{(0)}(\vec r,\vec r^\prime, E) \, = \,
-\,\frac{m_{red}}{4\,\pi\,\Gamma(1+i\,\rho)\,\Gamma(1-i\,\rho)}\,
\int\limits_0^1 \! dt \int\limits_1^\infty \! ds \,
\Big[\, s\,(1-t) \,\Big]^{i\,\rho}\,
\Big[\, t\,(s-1) \,\Big]^{- i\,\rho}\,\times
}
\nonumber\\[2mm] & &
\times\,\frac{\partial^2}{\partial t\,\partial s}\,
\bigg[\,
\frac{t\,s}{|\,s\,\vec r - t\,\vec r^\prime\,|}\,
\exp\bigg\{\,
i\,p\,\Big(\,
|\,\vec r^\prime\,|\,(1-t) + |\,\vec r\,|\,(s-1) +
|\,s\,\vec r - t\,\vec r^\prime\,|
\,\Big)
\,\bigg\}
\,\bigg]
\,,\quad r^\prime \, < \, r
\,,
\label{CoulombGreenfunctioncomplete}
\end{eqnarray}
where
\begin{equation}
p \, \equiv \, m_{red}\,v \, = \, \sqrt{m_{red} \, (E+i\,\epsilon)}
\,,\qquad
\rho \, \equiv \, \frac{C_F\,a_s}{2\,v}
\end{equation}
and $\Gamma$ is the gamma function. The case $r < r^\prime$ is
obtained by interchanging $r$ and $r^\prime$. $G_c^{(0)}(\vec
r,\vec r^\prime, E)$ represents the analytical expression for the
sum of ladder diagrams with Coulomb exchanges. The analytic form
of the Coulomb Green function shown in
Eq.~(\ref{CoulombGreenfunctioncomplete}) has been taken from
Ref.~\cite{Wichmann1}. Fortunately we do not need the Coulomb
Green function in its most general form but only its S-wave
component with one of the relative quark distances set to
zero.\footnote{In the section, discussing the nonperturbative
corrections, coming from gluon condensate operator, one more
representation of Coulomb Green function will be introduced.}
\begin{eqnarray}
G_c^{(0)}(0,r, E) & = & G_c^{(0)}(0,\vec r, E)
 \, = \,
-\,i\,\frac{m_{red}\,p}{2\,\pi}\,e^{i\,p\,r}\,
\int\limits_1^\infty\! dt \, e^{2\,i\,p\,r\,t}\,
\bigg(\,\frac{1+t}{t} \,\bigg)^{i\,\rho} \nonumber\\[2mm] & = &
-\,i\,\frac{m_{red}\,p}{2\,\pi}\,e^{i\,p\,r}\,
\Gamma(1-i\,\rho)\,U(1-i\,\rho,2,-2\,i\,p\,r)
\label{CoulombGreenfunctionzero}
\end{eqnarray}
where $U(a,b,z)$ is a confluent hypergeometric
function~\cite{Abramowitz1,Gradshteyn1}. It is an important fact
that $ G_c^{(0)}(0,\vec r, E)$ diverges for the limit $r\to 0$
because it contains power ($\propto 1/r$) and logarithmic
($\propto \ln r$) divergences~\cite{Hoang4}. As we have explained
before these ultraviolet (UV) divergences are regularized by
imposing the small distance cutoff $\mu_{\rm fact}$. The
regularized form of $\lim_{r\to 0}\,G_c(0,\vec r,E)$ reads
\begin{equation}
G_c^{(0),\,reg}(0,0,E) \, = \,
\frac{m_{red}^2}{4\,\pi}\,
\bigg\{\,
i\, v - C_F\,a_s\,\bigg[\,
\ln(-i \frac{m_{red}\,v}{\mu_{\rm fact}}) + \gamma_{\mbox{\tiny E}}
  + \Psi\bigg( 1-i\,\frac{C_F\,a_s}{2 v} \bigg)
\,\bigg]
\,\bigg\}
\,,
\label{CoulombGreenfunctionregularized}
\end{equation}
where the superscript ``reg'' indicates the cutoff regularization
and $\Psi(z)=d \ln\Gamma(z)/dz$ is the digamma function. For the
regularization we use the convention where all power divergences
$\propto \mu_{\rm fact}$ are freely dropped and only logarithmic
divergences $\propto \ln(\mu_{\rm fact}/m_{red})$ are kept.
Further, we define $\mu_{\rm fact}$ such that in the expression
between the brackets all constants except the Euler-Mascheroni
constant $\gamma_{\mbox{\tiny E}}$ are absorbed.  The results for
any other regularization scheme which suppressed power divergences
(like the $\overline{\mbox{MS}}$ scheme) can be obtained by
redefinition of the factorization scale. For convenience we
suppress the superscript ``reg'' from now in this work.

The Coulomb Green function contains $c\bar b$ bound state poles at the
energies $\sqrt{s}_n = m_b + m_c - C_F^2 a_s^2 m_{red}/4 n^2$
($n=1,2,\ldots\infty$). These poles come from the digamma function in
Eq.~(\ref{CoulombGreenfunctionregularized}) and correspond to the
nonrelativistic positronium state poles known from
QED~\cite{Braun1}. They are located entirely {\it below}
the threshold point $\sqrt{s}_{\mbox{\tiny thr}} = m_b + m_c$.
This can be seen explicitly from the expression for imaginary part of
Coulomb Green function $G_c^{(0)}(0,0,E)$
\begin{eqnarray}
\mbox{Im}\Big[\,
G_c^{(0)}(0,0,E)
\,\Big] &=&
4\,\pi\,m_{red}\,
\sum\limits_{n=1}^{\infty}|\Psi_n(0)|^2\,
       \delta(s-s_n) \, + \nonumber\\
&& \Theta(E)\,\frac{1}{4\pi}\,m_{red}^2\,\frac{C_F\,a_s\,\pi}
 {1-\exp(-\frac{C_F\,a_s\,\pi}{v})}\,,
\label{Rthreshnonrelativistic}
\end{eqnarray}
where $|\Psi_n(0)|^2 = (m_{red} C_F a_s)^3/8 \pi n^3$ is the
modulus squared of the LO nonrelativistic bound state wave
functions for the radial quantum number $n$. The continuum
contribution on the right-hand side of
Eq.~(\ref{Rthreshnonrelativistic}) is sometimes called
``Sommerfeld factor'' or ``Fermi factor'' in the literature. And
the second term from the first line in
Eq.~(\ref{Rthreshnonrelativistic}) describes the resonance
contributions. And finally, the corrections to the zero-distance
Coulomb Green function calculated below lead to higher order
contributions to the bound state energy levels poles and the
continuum.

Let us now come to the determination of the corrections to the
zero-distance Coulomb Green function coming from the remaining
term in the Schr\"odinger equation~(\ref{Schroed}). At
next-to-leading order only the one-loop contributions to the
Coulomb potential, $V_c^{(1)}$  have to considered. Using first
order TIPT in configuration space representation the NLO
corrections to $G_c^{(0)}(0,0,E)$ reads
\begin{eqnarray}
G_c^{(1)}(0,0,E) & = & -\,\int d^3\vec r \,
G_c^{(0)}(0,r,E)\,V_c^{(1)}(\vec r)\,G_c^{(0)}(r,0,E) \,.
\label{GreenfunctionNLO}
\end{eqnarray}
We will not calculate explicitly NLO correction to the Coulomb
Green function here as the goal of this paper is to calculate the
theoretical expressions for moments. The later can be most
conveniently calculated by dispersion integration, using the
following representation for the theoretical moments
\begin{eqnarray}
P_n^{th} = \frac{6}{\pi}\int_0^{\infty}\frac{ds}{s^n} \mbox{Im} \{
C_1(\mu_{hard},\mu_{soft}) G_c(0,0,E)\},
\end{eqnarray}
where $E = \sqrt{s}-m_b-m_c$.

\subsection{Determination of the short distance coefficients}

The short-distance coefficient $C_1$ and $C_2$ can be determined
by matching perturbative calculations of the matrix elements in
full QCD and NRQCD.  A convenient choice for matching is the
matrix element between the vacuum and the state $|c {\bar b}
\rangle$ consisting of a c and a ${\bar b}$ on their perturbative
mass shells with nonrelativistic four-momenta $p$ and $p'$ in the
center of momentum frame:  ${\bf p} + {\bf p}' = 0$.  The matching
condition is
\begin{equation}
\langle 0 | {\bar b} \gamma^0 \gamma_5 c
    | c {\bar b} \rangle \Bigg|_{\mbox{{\scriptsize QCD}}}
= C_1 \; \langle 0 | \chi_b^\dagger \psi_c
    | c {\bar b} \rangle \Bigg|_{\mbox{{\scriptsize NRQCD}}}
 \;+\; C_2 \; \langle 0 | ({\bf D} \chi_b)^\dagger \cdot
 {\bf D} \psi_c | c {\bar b}\rangle \Bigg|_{\mbox{{\scriptsize NRQCD}}}
\;+\; \ldots, \label{match}
\end{equation}

To determine the short distance coefficients to order $\alpha_s$,
we must calculate the matrix elements on both sides of
(\ref{match}) to order $\alpha_s$. It is sufficient to calculate
the order-$\alpha_s$ correction only for the coefficient $C_1$,
since as we already said the contribution proportional to $C_2$ is suppressed
by
$v^2$. The coefficient $C_1$ can be isolated by taking the limit
${\bf p} \to 0$, in which case the matrix element of $({\bf D}
\chi_b)^\dagger \cdot {\bf D} \psi_c$ vanishes. Such calculations
were done in \cite{Braaten}, where the following result was
obtained for $C_1$:
\begin{equation}
C_1 \;=\; 1 \;+\; {\alpha_s(m_{\rm red}) \over \pi}
    \left[{m_b-m_c \over m_b+m_c} \log {m_b \over m_c}
        - 2 \right],
\label{C0}
\end{equation}
where $m_{\rm red}$ is the scale of the running coupling constant.
The same result gives the direct matching of NRQCD correlator with
QCD one, taking into account factor 2 for $\alpha_s$ correction.

\section{Dispersion integration}

In this section we will discuss issues related to dispersion
integration in expressions for NRQCD moments. In general, the
integration of spectral density over complete covariant form of
integration measure $\frac{d s}{s^{n+1}}$ is quite cumbersome
However, in NRQCD approximation, as we will see soon, this task
significantly simplifies. Let us make a change of variables $E =
\sqrt{s}-m_b-m_c$  and consider a limit $E \ll m_b + m_c$. In this
case the integration measure takes the form
\begin{eqnarray}
\frac{d s}{s^{n+1}}&=& \frac{1}{(m_b+m_c)^{2n}}\frac{2 d E}{m_b+m_c}
\exp \{-(2n+1)\ln (1+\frac{E}{m_b+m_c})\} \\
&&\approx \frac{1}{(m_b+m_c)^{2n}}\frac{2 d E}{m_b+m_c}\exp \{\frac{2 E
n}{m_b+m_c}\} +
O(\frac{2 E}{m_b+m_c})  \nonumber
\end{eqnarray}
The dispersion integration for NRQCD moments in this limit is
\begin{equation}
P_n^{th} = \frac{1}{(m_b+m_c)^{2n}}\int_{E_{bind}}\frac{2 d E}{m_b+m_c}
\exp \{-\frac{2 E n}{m_b+m_c}\}R_{NLO}^{th}(E),
\end{equation}
where $E_{bind} = \frac{m_{red}C_F^2\alpha_s^2}{2}$ is the
negative binding energy of the lowest lying resonance. This
integration is performed most efficiently by deforming the path of
integration into negative complex energy plane, such that the part
of the integration path parallel to imaginary axis is far away
from bound state poles. That is
\begin{equation}
P_n^{th} = \frac{1}{(m_b+m_c)^{2n}}\frac{1}{2i}\int_{\gamma -
i\infty}^{\gamma + i\infty}\frac{2 d E}{m_b+m_c}\exp\{-\frac{2 E
n}{m_b+m_c}\}C_1 A_1 (E),\end{equation} where $\gamma \ll
E_{bind}$. Note, that in the above equation the real part of
correlator $A_1$ is also present, which is needed for integration
over the new path. After performing the second change of variables
$E\to - \tilde{E}$
\begin{equation}
P_{n}^{th} = \frac{\pi}{(m_b+m_c)^{2n}}\frac{1}{2\pi
i}\int_{\gamma - i\infty}^{\gamma + i\infty}\frac{2 d
\tilde{E}}{m_b+m_c}\exp\{\frac{2 \tilde{E} n}{m_b+m_c}\}C_1 A_1
(-\tilde{E})
\end{equation}
we see, that the problem of evaluation of NRQCD moments can be
related to the inverse Laplace transform of integrand expression,
for with there are a lot of tables in literature. The procedure of
taking inverse Laplace transform can be further simplified by
noting that integration path is far away from bound state energies
and hence the integrand can be safely expanded in $\alpha_s$.

Following the steps, described above, and using relations for
inverse Laplace transform from Appendix B the NRQCD moments in
leading order of NRQCD expansion have the form
\begin{equation}
[P_n^{th}]^{LO} = \frac{3
(m_{red})^{3/2}(m_b+m_c)^{1/2}}{2\sqrt{\pi}(m_b+m_c)^{2n}n^{3/2}}\Phi^0
(\gamma),
\end{equation}
where
\begin{equation}
\Phi^0 (\gamma) = 1 + 2\sqrt(\pi)\gamma + \frac{2\pi^2}{3}\gamma^2
+ 4\sqrt{\pi}\sum_{p=1}^{\infty}\left (\frac{\gamma }{p}\right )^3
\exp\{\left (\frac{\gamma}{p}\right )^2\}[1+\mbox{erf}\left
(\frac{\gamma}{p}\right )]
\end{equation}
and
\begin{equation}
\gamma\equiv \frac{C_F\alpha_s m_{red}^{1/2}
n^{1/2}}{(m_b+m_c)^{1/2}}
\end{equation}
The calculation of NLO order correction to moments, coming from
correction to potential (\ref{V1}) is a bit more involved. With
the help of presentation (\ref{CoulombGreenfunctionzero}) for
Coulomb Green function we have
\begin{eqnarray}
[P_n^{th}]^{NLO}
&=&\frac{6}{(m_b+m_c)^{2n}}\frac{1}{2i}\int_{\gamma -
i\infty}^{\gamma + i\infty}\frac{2 d
\tilde{E}}{m_b+m_c}\exp\{\frac{2 \tilde{E}
n}{m_b+m_c}\}\times\nonumber \\ &&\left \{4\pi\int_0^{\infty}r^2
dr\left (\frac{m_{red}k}{\pi}\right
)^2\int_0^{\infty}dt\int_0^{\infty}du e^{-2kr(t+u+1)}\left
(\frac{(1+t)(1+u)}{tu}\right )^{\lambda}\right \}\times\nonumber
\\ && C_F\left (\frac{\alpha_s^2}{4\pi}\right )\left \{2\beta_0 (\frac{1}{r}
\log (\tilde{\mu} r))+\frac{a_1}{r}\right \},
\end{eqnarray}
where $\lambda = \frac{C_F m_{red}\alpha_s}{k}$ and $k =
2m_{red}\tilde{E}$. The integration over $r$ can be easily
performed explicitly, while the situation with integrations over
$t$ and $u$ is far more complicated. However, as we noted above
the integrand expression in the case under consideration can be
expanded in series over $\lambda$\footnote{It's the same as the
expansion in $\alpha_s$} . Then the result of integration over all
terms in $\lambda$ expansion, except $\lambda^0$ can easily
expressed in terms of functions $w_p^1$ and $w_p^0$, introduced in
\cite{Hoang2}\footnote{The expressions for them can be found in
Appendix B}. As for the term with $\lambda^0$, which contain a
manifestly divergent integral, it can be easily calculated in
terms of $K (\tau) = \langle 0|\exp (-H\tau)|0\rangle$, using the
free evolution function $K_0 (\tau) = \langle 0|\exp
(-H_0\tau)|0\rangle$. Here $H = H_0 -\frac{C_F\alpha_s}{r}$, $H_0
= -\frac{\vec\nabla^2}{2m_{red}}$ and $\tau$ is an Euclidean time.
The evolution function $K (\tau )$ can be related to the $n$'th
NRQCD moment by the following relation
\begin{eqnarray}
P_n^{th} = \frac{6}{(m_b+m_c)^{2n}}\frac{2\pi}{m_b+m_c}
K(\frac{2n}{m_b+m_c})
\end{eqnarray}
Using this relation and explicit expression for free quark propagation
function:
\begin{equation}
\langle {\bf x}|\exp (-H_0\tau)|{\bf y}\rangle =
\left (\frac{m_{red}}{2\pi\tau}\right )^{3/2}\exp
\left (-\frac{m_{red}}{2\tau}(\bf{x}-\bf{y})^2\right ),
\end{equation}
we can now easily evaluate the contribution of
$\lambda^0$ term for NRQCD moments. The combined result for NLO
order correction to NRQCD moments reads
\begin{eqnarray}
[P_{n}^{th}]^{NLO} = \frac{3
(m_{red})^{3/2}(m_b+m_c)^{1/2}}{2\sqrt{\pi}(m_b+m_c)^{2n}n^{3/2}}\Phi^1
(\gamma),
\end{eqnarray}
where
\begin{eqnarray}
\Phi^1 (\gamma ) = \frac{2\beta_0\alpha_s}{\sqrt{\pi}}\gamma \left
\{ \frac{1}{2}\log
(\frac{\mu_1e^{\gamma_E}\sqrt{n}}{2\sqrt{m_{red}(m_b+m_c)}}) +
\sum_{p=1}^{\infty}\gamma^p [w_p^1+w_p^0 (\log \left ( \frac{2
m_{red}(m_b+m_c)}{\mu_1\sqrt{n}} \right )+\frac{1}{2}\Psi \left (
\frac{p}{2}\right ) )] \right\}
\end{eqnarray}
Here $\mu_1 = \mu_{soft}\exp \left (\frac{a_1}{2\beta_0} \right )$. The full
perturbative
result for NRQCD moments with account of hard gluon corrections reads
\begin{equation}
[P_n^{th}]^{pert} = (1 \;+\; {\alpha_s(m_{\rm red}) \over \pi}
    \left[{m_b-m_c \over m_b+m_c} \log {m_b \over m_c}
        - 2 \right])\{[P_n^{th}]^{LO} + [P_n^{th}]^{NLO}\}
\end{equation}

\section{Nonperturbative corrections}

In this subsection we would like to consider corrections, given by
gluon condesate operator. The calculation is very similar to the
one done previously by Voloshin and Leutwyler \cite{Voloshin} for
the case of equal quark masses. For the determination of
corrections to Coulomb Green function, coming from gluon
condensate operator, we will exploit the fact, that the size of
vacuum fluctuations of gluon field is much larger then the size of
$B_c$-meson state. So, we can perform a multipole expansion for
interaction of $B_c$-meson with gluon condensate, whose first term
is
\begin{equation}
H_{\rm int} = -\frac{1}{2}g\xi^a\vec r\vec E^a,
\end{equation}
where $\vec r =\vec x_{c} - \vec x_{\bar b}$, $g^2 =
4\pi\alpha_s$, $\xi^a = t_1^a - t_2^a$. By emploing colour and
Lorenz invariance of the vacuum state one can relate the average
over the vacuum state of two chromoelectric fields to the
manifestly Lorenz-invariant value
\begin{equation}
\langle 0|E_i^aE_k^b|0\rangle =
-\frac{1}{96}\delta_{ik}\delta^{ab} \langle 0|G_{\mu\nu}^c
G_{\mu\nu}^c|0\rangle
\end{equation}
Thus, the Coulomb Green function with account for gluon condesate
corrections has the following form
\begin{eqnarray}
G(\vec x,\vec y,E) &=& G_{(0)} - \frac{1}{18}\langle 0|\pi\alpha_s
G_{\mu\nu}^aG_{\mu\nu}^a|0\rangle\times\\ &&\int\int d^3\vec r
d^3\vec r'(\vec r\vec r')G_{(0)}(\vec x,\vec r,E) G_{(8)}(\vec
r,\vec r',E)G_{(0)}(\vec r',\vec y,E),\nonumber
\end{eqnarray}
where $G_{(0)}$ and $G_{(8)}$ are Coulomb Green functions in
singlet and octet colour states correspondingly. They are defined
as solutions to the following equation
\begin{equation}
\bigg(\frac{p^2}{2m_{red}} + V_{(0,8)}(|\vec x|) +
\frac{k^2}{2m_{red}}\bigg) G_{(0,8)}(\vec x,\vec
y,-\frac{k^2}{2m_{red}}) = \delta (\vec x - \vec y),
\end{equation}
where
\begin{equation}
V_{0}(r)=
-\frac{\alpha^{(0)}}{r}=-\frac{4}{3}\frac{\alpha_s}{r},\quad
V_{8}(r)= -\frac{\alpha^{(8)}}{r}=\frac{2}{3}\frac{\alpha_s}{r}.
\end{equation}

The expressions to NRQCD moments, coming form gluon condensate,
can be calculated with the use of wave decomposition of NRQCD
Green function
\begin{equation}
G_c^{(0,8)}({\vec x}, {\vec y}; E) =
\sum_{l=0}^{\infty}(2l+1)G_c^{(0,8)l}(x, y; E) P_l(({\vec x}{\vec
y})/xy),
\end{equation}
where
\begin{eqnarray}
G_c^{(0,8)l}(x, y; -k^2/2m_{red}) &=& \frac{m_{red}k}{\pi}(2kx)^l
(2ky)^le^{k(x+y)}\sum_{s=0}^{\infty}\frac{L_s^{2l+1}(2kx)L_s^{2l+1}(2ky)s!}
{(s+l+1-m_{red}\alpha^{(0,8)} /k)(s+l+1)!},\nonumber \\
\end{eqnarray}
where $x = |{\vec x}|$, $y = |{\vec y}|$. $P_l$ and $L_s^p$ are
Legendre and Laguerre polynomials. The result for gluon condensate
correction to Coulomb Green function is\footnote{Here we assume
that one should perform the dispertion integration in order to
obtain the gluon condensate correction to NRQCD moments}:
\begin{equation}
[P_n^{th}]^{G^2} = -\frac{\sqrt{\pi
}(m_{red})^{1/2}(m_b+m_c)^{1/2}n^{3/2}} {12(m_b+m_c)^{2n+3}}\chi
(\gamma)\langle 0|\alpha_sG_{\mu\nu}^aG_{\mu\nu}^a|0\rangle
\end{equation}

The function $\chi (\gamma )$ is given by the following equation
\begin{eqnarray}
\chi (\gamma ) &=&
4\sqrt{\pi}\{\sum_{p=1}^{\infty}(p+1)(p+2)(p+3)\{\frac{1}{9p+16}[
\phi_2 (\frac{\gamma}{p}) - \phi_1
(\frac{\gamma}{p})(1+\frac{p}{12}(25-\frac{6}{9p+16}))] \nonumber
\\ && + \frac{16}{9p+17}[\phi_2 (\frac{\gamma}{p+1}) - \phi_1
(\frac{\gamma}{p+1})(1+\frac{p+1}{6}(5-\frac{3}{9n+17}))]+ \\ &&
\frac{4}{p+2}[\phi_2 (\frac{\gamma}{p+2}) - \frac{17}{18}\phi_1
(\frac{\gamma}{p+2})] + \frac{16}{9p+19}[\phi_2
(\frac{\gamma}{p+3}) - \phi_1 (\frac{\gamma}{p+3})(1-\nonumber \\
&& \frac{p+3}{6}(5+\frac{3}{9p+19}))] + \frac{1}{9p+20}[\phi_2
(\frac{\gamma}{p+4}) - \phi_1
(\frac{\gamma}{p+4})(1-\frac{p+4}{12}(25+\nonumber \\ &&
\frac{6}{9p+20}))]\} +
\frac{4}{81}\sum_{p=2}^{\infty}\frac{p^2-1}{(81p^2-1)(81p^2-4)}
\phi_1 (-\frac{\gamma}{8p})\},\nonumber
\end{eqnarray}
where
\begin{eqnarray}
\phi_1 (x) &=&
x^{-3}[e^{x^2}(1+\mbox{erf}(x))-1]-(\frac{2}{\sqrt{\pi}})x^{-2} -
x^{-1}, \\ \phi_2 (x) &=&
[e^{x^2}(1+\mbox{erf}(x))-1]x^{-1}-\phi_1 (x).
\end{eqnarray}
The above expression in the limit of equal quark masses coincides
can be checked with the one derived previously by M.B.Voloshin
\cite{Voloshin}. However, this expression is quite complicated and
in the range $\gamma \le 1.5$ it is far more convenient to use an
approximated formulum $\chi (\gamma ) = e^{-0.80\gamma}\Phi^0
(\gamma )$.

\section{Numerical results}

\begin{center}
\begin{figure}[ph]
\vspace*{-1. cm} \hbox to 1.5cm
{\hspace*{3.cm}\hfil\mbox{$\frac{M_n^{exp}}{M_n^{th}}$}}
\vspace*{7.5cm} \hbox to 15.cm {\hfil \mbox{$n$}}
\vspace*{7.5cm}\hspace*{2.5cm} \epsfxsize=12cm \epsfbox{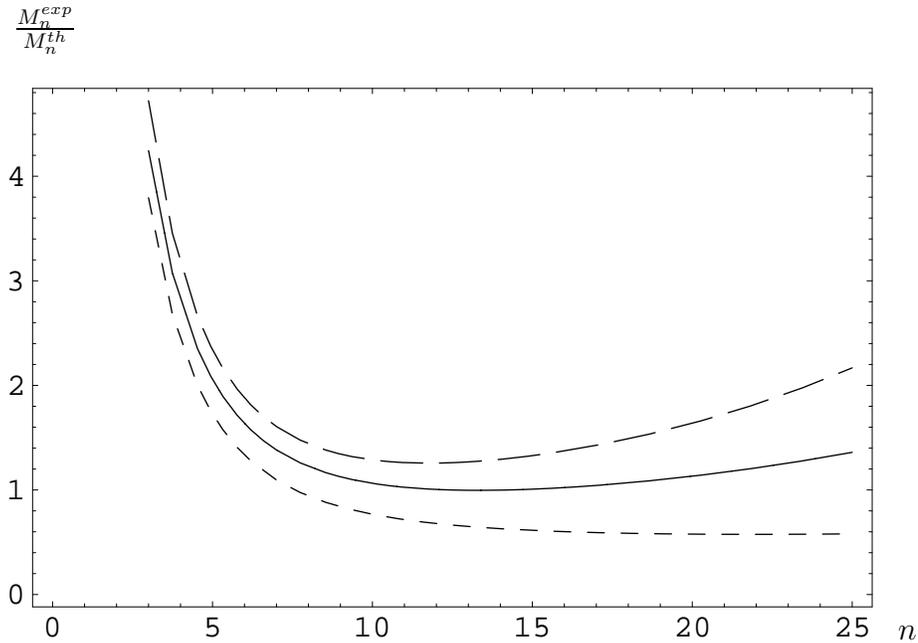}
\vspace*{-15.cm} \caption{ Ratios of experimental and
theoretically calculated moments $\frac{M_n^{exp}}{M_n^{th}}$ as
functions of momentum number (The solid line is for $M_c^{pole} =
2.03$ GeV, the line with short dashes is for $M_c^{pole} = 1.9$
GeV  and  the one with long dashes is for $M_c^{pole} = 2.1$ GeV.
All ratios are plotted for $\mu_{hard} = 2.0$ GeV and $\mu_{soft}
= 1.1$ GeV ).} \label{fig1}
\end{figure}
\end{center}

In this section we will discuss our numerical estimates for
various $c$-quark mass definitions, the $B_c$-meson mass and
coupling constant. Let us first discuss our anzaz for experimental
spectral density, the need for which is dictated by our present
lack of experimental measurements on higher $B_c$-meson states.
The experience in potential models and quasi-local sum rules \cite{Kis2} allow
us to write
\begin{eqnarray}
M_n &=& M_1+2T\ln n,\quad M_1=6.3\;{\rm GeV},\quad T=0.415\;{\rm GeV},
\nonumber \\
\frac{f_n^2}{f_l^2} &=& \frac{1}{n}\frac{M_l}{M_n},\quad\;\;\;
f_1(0^{-})=0.4\;{\rm GeV},
\end{eqnarray}
where $M_n$ is the mass of $nS$ $B_c$-meson state, and $f_n$ is
its coupling. So that, the experimental spectral density has the
form
\begin{eqnarray}
\rho (s)_{hard} = \pi\sum_{n=1}^{3}\frac{f_n^2
M_n^4}{(m_b+m_c)^2}\delta (s - M_n^2) + \rho (s)_{pert}\theta (s -
s_{thr} )
\end{eqnarray}
Now with the knowledge of experimental spectral density and thus
of experimental moments we can compare them with theoretical ones.
To begin with let's see what values of the pole $c$-quark mass
will give us the required values of experimental moments. For the
value of $b$-quark pole mass $M_b^{pole}$, needed for calculation
of theoretical moments\footnote{The precise values of other
parameters, that is gluon condensate and continuum threshold are
not so important, as the correlation function depends on them
weakly.}, we have taken the result of NNLO analysis for $\Upsilon$
family\cite{NNLO}, so that we have $M_b^{pole} = 4.8\pm 0.06$ GeV.
Varying the soft scale in $\mu_{soft}$ the range $1.1 - 1.2$ GeV
and fixing\footnote{Note, that at NLO, due to vanishing of
anomalous dimension of the current under consideration, the
dependence of two-point correlator  from $\mu_{hard}$ enters only
through $\alpha_s (\mu_{hard})$. So, the particular value of
$\mu_{hard}$ is not very important.} the hard scale $\mu_{hard}$
to $2.$ GeV, from the stability of the ratios of experimental to
theoretical moments we have that $M_c^{pole} = 1.7 - 2.1$ GeV. The
range of pole $c$-quark mass obtained is quite large and to reduce
it we need some extra condition to satisfy. Again, from the
spectroscopy of $B_c$-meson family it is known that the
characteristic value of strong coupling constant for heavy quark
dynamics inside $B_c$-meson is in the range $\alpha_s =
0.43-0.48$. This extra requirement heavily constrains the value of
$c$-quark pole mass, so that $M_c^{pole} = 2.03\pm 0.06$ GeV. At
Fig. \ref{fig1} we have shown the ratios of experimental to
theoretical moments as functions of momentum number at fixed value
of $\mu_{soft} = 1.1$ GeV and different values of $c$-quark pole
mass.

\vspace*{0.5cm}
The performed analysis shows that to satisfy chosen criteria we
need very low values of $\mu_{soft}\approx 1.1$ GeV or lower. On
the other hand, as was shown in \cite{Kis4}, to match perturbative
heavy quark potential with full QCD potential at characteristic
distances of quarks bound inside heavy-heavy mesons, we should
have the soft scale $\mu_{soft}$ in the region $1.3-2.0$
GeV\footnote{Such choice of soft scale is also desirable in order
to have reliable perturbative predictions for theoretical
moments.} To achieve this goal we modify the original TIPT for
theoretical moments by shifting the leading order Coulomb
potential by a part of constant term in NLO potential, that is
\begin{eqnarray}
V^{(0)}_{modified} = -\frac{C_F\alpha_v}{r} =
-\frac{C_F\alpha_s}{r}(1 + t\left(\frac{\alpha_s}{4\pi}\right))
\end{eqnarray}
The resulting theoretical moments will now depend not only from
hard and soft scales but also from the value of the shift
parameter $t$. To get rid of the latter dependence, while keeping
in mind our goal, we perform optimization in parameter $t$,
require the NLO corrections to theoretical moments at given
momentum number $n$ do not exceed $1 \%$\footnote{Note, that here
we will need already $\alpha_v(\mu_{soft})= 0.43-0.48$}. As a
result we get $M_c^{pole} = 1.96\pm 0.05$ GeV. At Fig. \ref{fig1}
we show the ratios of experimental to theoretical moments as
functions of momentum number at fixed value of $\mu_{soft} = 1.3$
GeV and different values of $c$-quark pole mass. We see, that
within the error bars our results for the $c$-quark pole mass from
these two estimates agree with each other as well as with the
results of other estimates \cite{Kis4}.

\vspace*{0.5cm}
\begin{center}
\begin{figure}[ph]
\vspace*{-1. cm} \hbox to 1.5cm
{\hspace*{3.cm}\hfil\mbox{$\frac{M_n^{exp}}{M_n^{th}}$}}
\vspace*{7.5cm} \hbox to 15.cm {\hfil \mbox{$n$}}
\vspace*{7.5cm}\hspace*{2.5cm} \epsfxsize=12cm \epsfbox{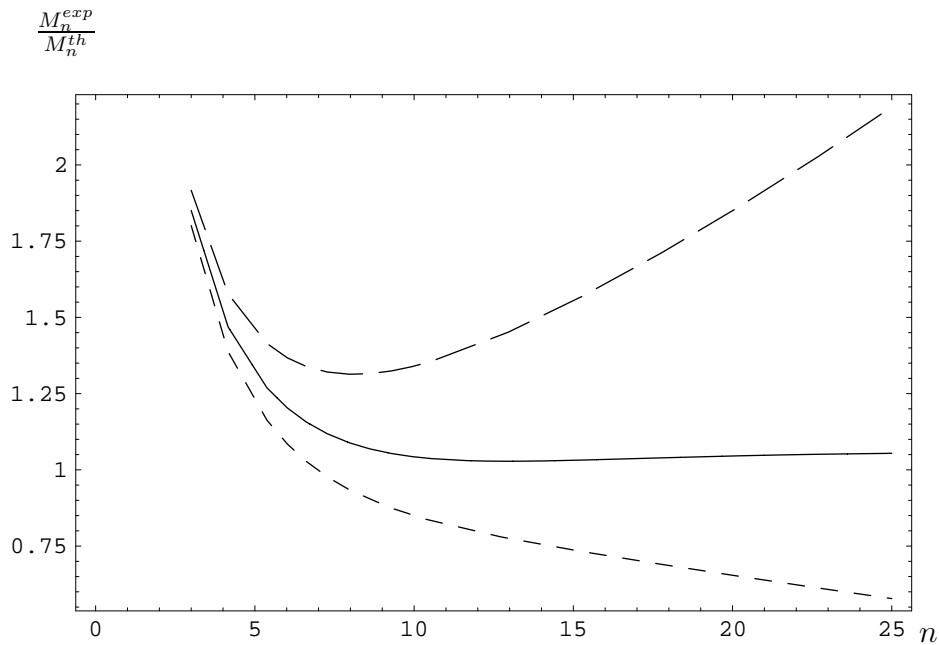}
\vspace*{-15.cm} \caption{ Ratios of experimental and
theoretically calculated moments $\frac{M_n^{exp}}{M_n^{th}}$ as
functions of momentum number in modified TITP (The solid line is
for $M_c^{pole} = 1.96$ GeV, the line with short dashes is for
$M_c^{pole} = 1.85$ GeV  and  the one with long dashes is for
$M_c^{pole} = 2.05$ GeV. All ratios are plotted for $\mu_{hard} =
2.0$ GeV and $\mu_{soft} = 1.3$ GeV ).} \label{fig2}
\end{figure}
\end{center}

\begin{center}
\begin{figure}[ph]
\vspace*{-1. cm} \hbox to 1.5cm
{\hspace*{3.cm}\hfil\mbox{$\frac{M_n^{exp}}{M_n^{th}}$}}
\vspace*{7.5cm} \hbox to 15.cm {\hfil \mbox{$n$}}
\vspace*{7.5cm}\hspace*{2.5cm} \epsfxsize=12cm \epsfbox{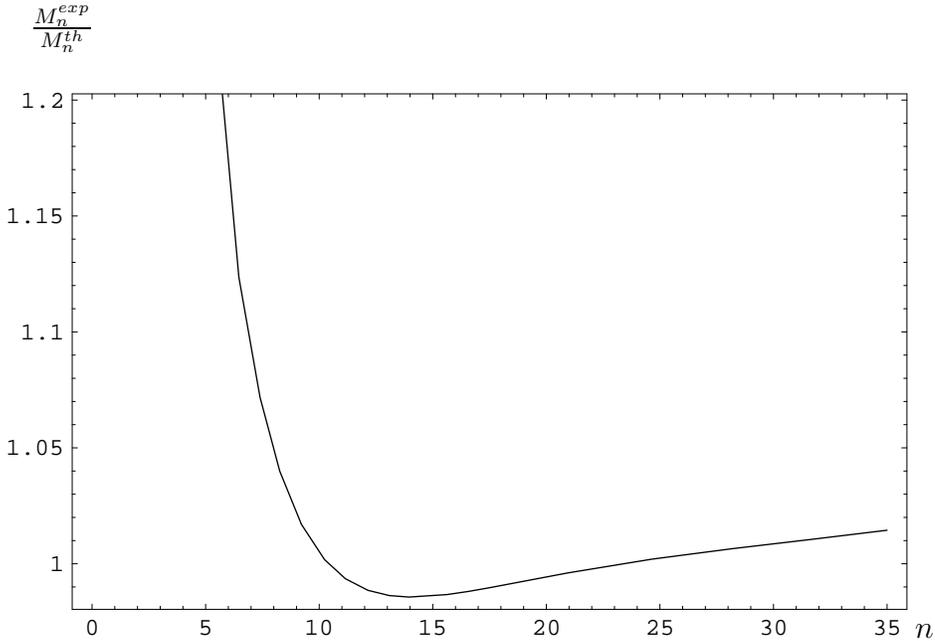}
\vspace*{-15.cm} \caption{ Ratio of experimental and theoretically
calculated moments $\frac{M_n^{exp}}{M_n^{th}}$ as functions of
momentum number in modified TITP ($M_c^{1S} = 1.52$ GeV,
$\mu_{hard} = 2.0$ GeV and $\mu_{soft} = 1.3$ GeV ).} \label{fig3}
\end{figure}
\end{center}

Having extracted the value of $c$-quark pole mass from $B_c$-sum
rules, an analogous analysis can be performed for other
definitions of heavy quark masses. Below we give the numerical
values of these masses obtained in the same way and comment on
their relation with each other. First, let us consider the running
$c$-quark mass related to the pole one by the following relation
\cite{Chet,Mel}
\begin{eqnarray}
\frac{M^{pole}}{\bar m(\bar m)} &=& 1 + 1.333
\left(\frac{\alpha_s(\bar m )}{\pi}\right) + [13.44-1.041
n_l]\left(\frac{\alpha_s(\bar m )}{\pi}\right)^2 \nonumber \\ && +
[190.1 - 26.7 n_l + 0.653 n_l^2]\left(\frac{\alpha_s(\bar m
)}{\pi}\right)^3
\end{eqnarray}
Our estimates for the $\bar m(\bar m)$ mass are
\begin{equation}
\bar m_c (\bar m_c) = 1.40\pm 0.07~{\mbox GeV}
\end{equation}
However, not pole not running quark masses are qood ones when one
works in the region of energies close to threshold. As we have
seen in the analysis performed above the two-point correlation
function has a strong dependence on renormalization scales as well
as correlation between the pole mass and strong coupling constant
(the situation is similar for running quark masses), so to reduce
this dependence somewhat it's would be more appropriate to explore
quark masses, whose use can eliminate as much as possible all such
dependencies and correlations among parameters. For this reason
here we give estimates of $1S$ and potential subtracted $c$-quark
masses. The former can be related to the pole mass $M^{pole}$ with
the help of following formulae \cite{Hoang6}
\begin{eqnarray}
M^{1S} &=& M^{pole}[1 - \triangle^{LO} - \triangle^{LO}\delta^1 -
\triangle^{LO}\delta^2]
\end{eqnarray}
where
\begin{eqnarray}
\triangle^{LO} &=& \frac{C_F^2\alpha_s^2}{8},\\ \delta^1 &=&
\left(\frac{\alpha_s}{\pi}\right)[\beta_0(L + 1) + \frac{a_1}{2}],
\\ \delta^2 &=&
\left(\frac{\alpha_s}{\pi}\right)^2[\beta_0^2(\frac{3}{4}L^2 + L +
\frac{\zeta_3}{2} + \frac{\pi^2}{24} + \frac{1}{4}) +
\beta_0\frac{a_1}{2}(\frac{3}{2}L + 1) + \frac{\beta_1}{4}(L + 1)
\nonumber \\ && + \frac{a_1^2}{16} + \frac{a_2}{8} + (C_A -
\frac{C_F}{48})C_F\pi^2], \\ L &\equiv& \ln (\frac{\mu
}{C_F\alpha_sM^{pole}}),
\end{eqnarray}
and
\begin{eqnarray}
a_2 &=& \left(\frac{4343}{162} + 4\pi^2 - \frac{\pi^4}{4} +
\frac{22}{3}\zeta_3\right)C_A^2 - \left(\frac{1798}{81} +
\frac{56}{3}\zeta_3\right)C_A T n_l \nonumber \\ && -
\left(\frac{55}{3} - 16\zeta_3\right)C_F T n_l +
\left(\frac{20}{9}T n_l\right)^2
\end{eqnarray}
With the use of these formulae we have the following estimate for
$1S$ $c$-quark mass at optimized value of $\mu_{soft} = 1.3$
GeV\footnote{This value is taken from the previous analysis for
$c$-quark pole mass performed in modified TIPT, as the shift of
leading order Coulomb potential implicitly includes part of NNLO
corrections. However, as we do not have up to date the complete
NNLO order analysis this result should be taken with care.}
\begin{equation}
M^{1S} = 1.49\pm 0.07~\mbox{GeV}
\end{equation}
Having done this estimate it is instructive to compare it with the
estimates of NLO analysis done in 1S scheme. To obtain the
expressions for theoretical moments in this scheme we make a
following substitution
\begin{eqnarray}
&&\frac{1}{(m_b+m_c)^{2n}}\to \nonumber \\
&&\frac{1}{(M_b^{1S}+M_c^{1S})^{2n}}\exp^{-2n\triangle^{LO}(\alpha_s)}
\{1-2n\frac{M_b^{1S}}{M_b^{1S}+M_c^{1S}}\triangle^{LO}(\alpha_s)
\delta^1(M_b^{1S},\alpha_s,\mu_{soft})-\nonumber \\ &&
2n\frac{M_c^{1S}}{M_b^{1S}+M_c^{1S}} \triangle^{LO}(\alpha_s)
\delta^1(M_c^{1S},\alpha_s,\mu_{soft}) \}
\end{eqnarray}
In all other places the heavy quark pole masses should be changed
to $1S$ masses. Varying the scales in the same ranges as we did
for pole $c$-quark mass we obtain, that the sum rules are stable
themselves, but the ratio of experimental to theoretical moments
is far from unity for reasonable values of $1S$ $c$-quark
mass\footnote{Recall that this mass by definition differs from a
half of $J/\Psi$ mass on a small value given by nonperturbative
corrections.}. To solve this problem we again consider the
modified TIPT for theoretical moments. It is important, when
writing expressions for theoretical moments, to make for
consistency a similar shift in NLO order relation between pole and
$1S$ masses. Performing calculations in modified $1S$ scheme we
obtain the following estimate for $1S$ $c$-quark mass: $M_c^{1S} =
1.52\pm 0.05$ GeV, which, within the errors, is in agreement with
previous estimate for $1S$ $c$-quark mass. Fig. \ref{fig3} shows
us the ratios of experimental to theoretical moments as functions
of momentum number at fixed values of $\mu_{soft} = 1.3$ GeV and
$1S$ $c$-quark mass.

\vspace*{0.5cm}
To finish the discussion of $1S$ $c$-quark mass we note, that even
this mass was defined as $M_c^{pole}$ minus one half of
perturbative Coulomb energy in $J/\Psi$-meson, it can be equally
well applied for $B_c$-meson as well as other systems, containing
$c$-quark. The arguments here are the analog of Upsilon expansion
for charmed quarks and the fact, that the Coulomb energy for
static quarks does not depend on their flavors.

The estimate for the value of potential subtracted $c$-quark mass
was obtained from the relation of the latter to the running quark
mass
\begin{eqnarray}
m_{PS}(\mu_f) &=& \bar m(\bar m)\{1 + \frac{4\alpha_s(\bar
m)}{3\pi}[1-\frac{\mu_f}{\bar m(\bar m)}] +
\left(\frac{\alpha_s(\bar m )}{\pi}\right)^2[13.44-1.04 n_l
\nonumber \\ &&-\frac{\mu_f}{3\bar m(\bar
m)}(a_1+4\pi\beta_0[\ln\frac{\mu_f^2}{(M^{pole})^2}-2])]\}
\end{eqnarray}
Thus, at $\mu_f = 1.5$ GeV we have
\begin{equation}
m_{PS}(1.5 \mbox{GeV}) = 1.42\pm 0.07 \mbox{GeV}
\end{equation}
Having completed the estimates of different $c$-quark masses, we
can now perform estimates of the mass and coupling constant for
the ground $B_c$-meson state, fixing the heavy quark masses at
their central values.

\vspace*{0.5cm}
\begin{center}
\begin{figure}[ph]
\vspace*{-1. cm} \hbox to 1.5cm
{\hspace*{3.cm}\hfil\mbox{$M_{B_c}$}} \vspace*{7.cm} \hbox to
15.cm {\hfil \mbox{$n$}} \vspace*{7.5cm}\hspace*{2.5cm}
\epsfxsize=12cm \epsfbox{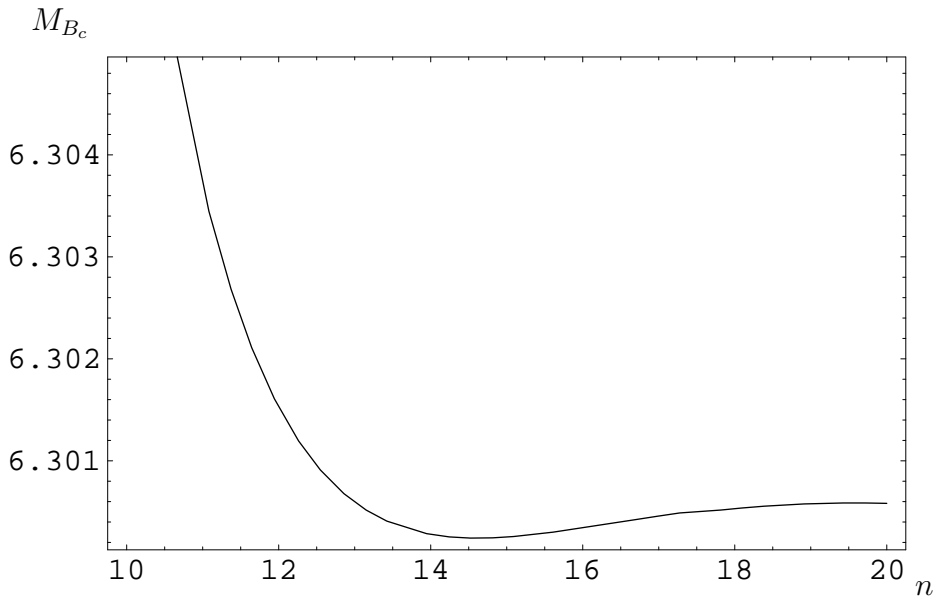} \vspace*{-15.cm} \caption{ The
values of $B_c$ - meson mass extracted from the two-point NRQCD
sum rules in moment scheme.} \label{fig4}
\end{figure}
\end{center}

\begin{center}
\begin{figure}[th]
\vspace*{-1. cm} \hbox to 1.5cm
{\hspace*{3.cm}\hfil\mbox{$f_{B_c}$}} \vspace*{7.cm} \hbox to
15.cm {\hfil \mbox{$n$}} \vspace*{7.5cm}\hspace*{2.5cm}
\epsfxsize=12cm \epsfbox{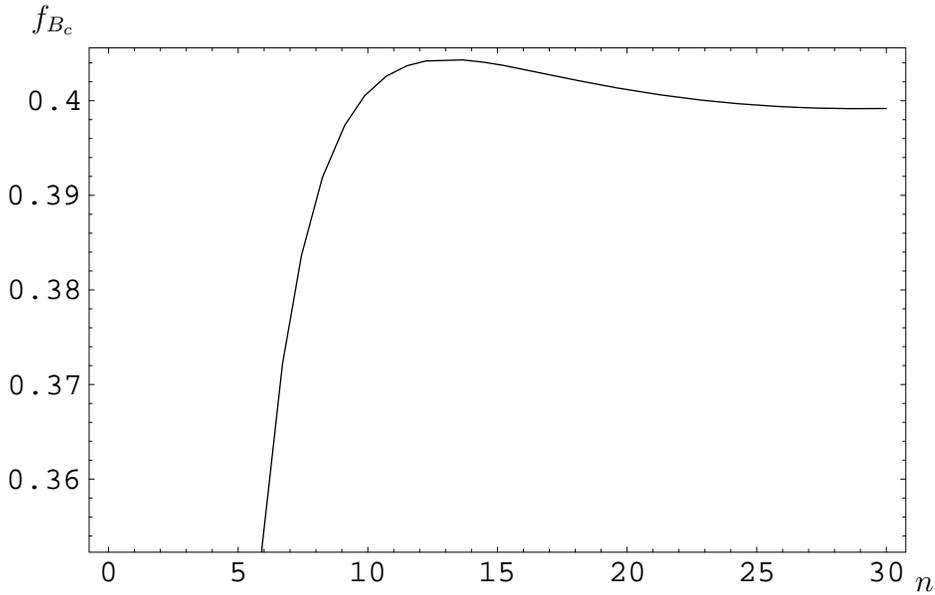} \vspace*{-15.cm} \caption{ The
values of $B_c$ - meson coupling constant extracted from the
two-point NRQCD sum rules in moment scheme.} \label{fig5}
\end{figure}
\end{center}

The results for mass and coupling of $B_c$-meson in momentum
scheme of NRQCD sum rules can be easily seen from Fig. \ref{fig4}
and \ref{fig5}.

\vspace*{0.5cm}

Recently, the estimate of $B_c$ mass was obtained in the perturbative potential
approach by fitting the masses of $J/\Psi$ and $\Upsilon$ in order to get a
good covergency in $\alpha_s$ and extract the heavy quark masses \cite{vairo}.
The perturbative mass $m_{B_c} = 6.323\pm 0.007$ GeV is very close to the our
estimate in the framework of QCD sum rules. This fact indicates a small
nonperturbative correction, which was limited by $-60$ MeV in \cite{vairo}.

\section{Conclusion}
In this paper we have presented complete NLO analysis for
$B_c$-meson two-point NRQCD sum rules. Analitycal results for
perturbative spectral density and gluon condensate contribution
with account for summed Coulomb corrections are derived and
analyzed. A detail numerical analysis as well as discussion on the
determination of various $c$-quark masses together with couplings
and masses of lowest lying $B_c$-meson resonances from the
mentioned sum rules are provided. The analysis shows that to
reduce the uncertainties of calculations it is mandatory to have
complete NNLO expressions for theoretical moments, which we plan
to accomplish in nearest future.

The author is grateful to Prof. V.V.Kiselev for stimulating
discussions and comments. I especially thank my wife for strong
moral support and help in doing physics.

This work was in part supported by the Russian Foundation fro
Basic Research, grants 99-02-16558 and 00-15-96645, by
International Center of Fundamental Physics in Moscow,
International Science Foundation and INTAS-RFBR-95I1300.

\appendix
\section*{Appendix A}

In this Appendix we present the two-loop result for perturbative spectral
density of
$B_c$ meson \cite{Broadh1}:
\begin{eqnarray}
\rho_{\rm pert} (t) &=& \frac{3}{8\pi t}\bar q^4 v\bigg\{1 +
\frac{4\alpha_s}{3\pi}\bigg\{\frac{3}{8}(7-v^2)\nonumber \\
&& + \sum_{i=b,c}\bigg[(v+v^{-1})(L_2(\alpha_1\alpha_2)) - L_2(-\alpha_i) -
\log\alpha_1\log\beta_i \\
&& A_i\log\alpha_i + B_i\log\beta_i\bigg]\bigg\} +
O(\alpha_s^2)\bigg\}\nonumber
\end{eqnarray}
where
\begin{equation}
L_2 (x) = -\int_0^{x}\frac{d y}{y}\log (1-y)
\end{equation}
and
\begin{eqnarray}
A_i &=& \frac{3}{2}\frac{3m_i + m_j}{m_i + m_j} - \frac{19+2v^2+3v^4}{32v} -
\frac{m_i (m_i - m_j)}{\bar q^2 v (1+v)}\bigg(
1+v+\frac{2v}{1+\alpha_i}\bigg);\nonumber\\
B_i &=& 2+2\frac{m_i^2-m_j^2}{\bar q^2v};\\
\alpha_i &=& \frac{m_i}{m_j}\frac{1-v}{1+v};\quad \beta_i = \sqrt{1+\alpha_i}
\frac{(1+v)^2}{4v}\nonumber\\
\bar q^2 &=& t - (m_b -m_c)^2;\quad v = \sqrt{1-4\frac{m_bm_c}{\bar q^2}}
\end{eqnarray}

\section*{Appendix B}

In this Appendix we have collected some formulae, needed for calculation of
NRQCD moments
in next to leading order.
\begin{eqnarray}
\frac{1}{2\pi i}\int_{\gamma - i\infty}^{\gamma +
i\infty}\frac{1}{x^{\nu}}e^{xt}dx &=&
\frac{t^{\nu -1}}{\Gamma (\nu)}, \\
\frac{1}{2\pi i}\int_{\gamma - i\infty}^{\gamma + i\infty}\frac{\ln
x}{x^{\nu}}e^{xt}dx &=&
\frac{t^{\nu -1}}{\Gamma (\nu)}[\Psi (\nu )-\ln t],
\end{eqnarray}
\begin{eqnarray}
w_p^0 &=& -\frac{1}{p!\Gamma (\frac{p}{2})}\int_0^{\infty}dt\int_0^{\infty}du
\frac{1}{(1+t+u)^2}\ln^p\left (\frac{(1+t)(1+u)}{tu}\right ) =
-\frac{(p+1)\zeta_{p+1}}
{\Gamma (\frac{p}{2})}, \\
w_p^1 &=& \frac{1}{p!\Gamma (\frac{p}{2})}\int_0^{\infty}dt\int_0^{\infty}du
\frac{1-\ln (1+t+u)}{(1+t+u)^2}\ln^p\left (\frac{(1+t)(1+u)}{tu}\right
)\nonumber \\
&=& -\left\{ \frac{(1+p)}{\Gamma (\frac{p}{2})}\left [\gamma_E\zeta_{p+1} +
\sum_{m=0}^{\infty}\frac{\Psi (2+m)}{(1+m)^{p+1}}\right ]+
\frac{2}{\Gamma (\frac{p}{2})}\sum_{l=0}^{p-1}\sum_{m=0}^{\infty}(-1)^{p-l}
\frac{(1+l)\Psi^{(p-l)}(2+m)}{(p-l)!(1+m)^{1+l}}\right \},\nonumber \\
\end{eqnarray}
where
$\zeta_p$ is the Rieman zeta function for argument $p$.
\end{document}